\documentstyle[aps,psfig]{revtex}
\begin{document}
\draft
\bibliographystyle{prsty}
\twocolumn[\hsize\textwidth\columnwidth\hsize\csname@twocolumnfalse\endcsname
%
\title{Extremal Coupled Map Lattices}
\author{ Guillermo Abramson$^{\ast}$ and Jos\'e Luis Vega$^{\dagger}$}
\address{Max-Planck-Insitut f\"ur Physik komplexer Systeme,
N\"othnitzer Str. 38, 01187 Dresden, Germany}
\author{\emph{(to appear in The European Physical Journal B, in press, 1999)}}
\date{\today}
\maketitle

\widetext
\begin{abstract}
We propose a model for co-evolving ecosystems that takes into
account two levels of description of an organism, for instance
genotype and phenotype. Performance at the macroscopic level
forces mutations at the microscopic level. These, in turn,
affect the dynamics of the macroscopic variables. In some
regions of parameter space, the system self-organises into a
state with localised activity and power law distributions.
\end{abstract}
\pacs{PACS: 05.45.+b, 64.60.Lx, 87.10.+e}

\vspace{1cm}
\vfill
] 
\narrowtext 

Complex extended systems showing  a lack of scale in their
features appear to be widespread in nature, being as diverse as
earthquakes~\cite{CJ89,KS95}, creep phenomena~\cite{Zaitsev92},
material fracturing~\cite{PPVAC94,DMP91,CTP96}, fluid
displacement in porous media~\cite{WW83,CR88}, interface growth
\cite{Sneppen92,Sneppen93}, river
networks~\cite{RRRIB93,MCFCB96,CGMRR96} and  biological
evolution~\cite{Raup86,EG72,GE77,EG88}.   At variance with
equilibrium statistical mechanics, these systems do not need any
fine tuning of a parameter to be in a critical state. To explain
this behaviour, Bak, Tang and Wiesenfeld introduced the concept
of self-organised criticality (SOC) through the simple sand-pile
model~\cite{BTW87,BTW88}. In recent years, several models with
extremal dynamics have been shown to exhibit SOC in the presence
of a sufficiently decorrelated signal ~\cite{PMB96}.

Several years ago, Bak and Sneppen (BS) proposed a SOC model
\cite{BS93} for the co-evolution of species. In the BS model
each species occupies a site on a lattice. Each  site $j$ is
assigned a fitness, namely a random number between $0$ and $1$.
At each time step in the simulation the smallest fitness is
found.  Then the fitnesses of the minimum and of its nearest
neighbours are  updated   according to the rule
\begin{equation}
f_{n+1}=F(f_{n})
\label{eq:noise}
\end{equation}
that assigns a new fitness $f_{n+1}$ at time $n+1$ to the chosen
lattice site. This corresponds to the extinction of the less fit
and its impact on the ecosystem. Indeed, in the original BS
model, the function $F$ is just a random function with a uniform
distribution between 0 and 1. The system reaches a stationary
critical state in which the distribution of fitnesses is zero
below a certain  threshold and uniform above it (the actual
value of the threshold depends on  the updating rule). It has
been shown \cite{DVV97,DVV98} that the exact  nature of the
updating rule is not relevant. Indeed,  the use of a chaotic map
instead of a random update, preserves the universality class
(even if the final distribution may be altered).

As a result of the  dynamical rules, the BS system exhibits
sequences of causally connected evolutionary events called {\em
avalanches} \cite{BS93}. The number of avalanches $N$ follows a
power law distribution $N(s)\sim s^{-\tau}$ where $s$ is the
size of the avalanche  and $\tau\sim 1.07$
\cite{PMB96,MPB94,Grassberger95}. Other  quantities also exhibit
a power law behaviour  with their own critical exponents.
Prominent among them are the first and all return time
distributions of activity (a site is defined as active when its
fitness is the minimum one),
\begin{equation}
P_f(t)\sim t^{-\tau_f}, \,\,\,\,P_a(t)\sim t^{-\tau_a}\, ,
\label{return}
\end{equation}
where $\tau_f\sim 1.58$ and $\tau_a\sim 0.42$
\cite{PMB96,MPB94,Grassberger95}.

In the BS model,  two basic ingredients
are needed for SOC to occur \cite{DVV98}:\\
i) Order from extremal dynamics (minimum rule).\\
ii) Disorder from the updating rule (stochastic or otherwise).

An oversimplification in the BS model is apparent: each species
is described by a single variable. The minimum rule is applied
to this variable, and so is the effect of mutations. In natural
evolving systems, however, at least two---interacting---levels
of organisation are present, and both play a role in the
evolution. Mutations occur at a microscopic level, namely at the
molecular level of the genotype. This affects a macroscopic
level, defining the phenotype. Natural selection acts on the
phenotype, allowing the survival of some and the extinction of
others, according to the observed power law distributions.

In this paper we propose a new model, namely an Extremal Coupled
Map Lattice (ECML), that takes into account this two level
structure. The ecosystem is represented as an ensemble of $N$
species arranged  on a one-dimensional lattice. Each species is
described by means of macroscopic variable $x_i$ subject to a
nonlinear dynamics and coupled to its nearest neighbours. In
general, $x_i$ can be identified with a population, or some
other function of the phenotype. The control parameter
$\lambda_i$ of the nonlinear dynamics is identified with the
microscopic level (genotype). To find the exact function
connecting these two levels of description is beyond the scope
of this simplified model. For this reason, we chose a logistic
map for the independent evolution of $x_i$ with $\lambda_i$
acting as the nonlinear parameter in the map \cite{comment2}.
With this in mind, the evolution of each species is given by:
\begin{equation}
x^i_{n+1} =
(1-\epsilon)f(x^i_n)+
\frac{\epsilon}{2}\left[f(x^{i-1}_n)+f(x^{i+1}_n)\right]
\label{cml}
\end{equation}
where $f(x)$ is the logistic map $f(x) = \lambda x(1-x)$ (in
general any chaotic function should do). Each site has its  own
$\lambda_i$,  extracted from a fixed distribution $g(\lambda)$
\cite{comment1}. The local coupling of strength $\epsilon_i$
emulates the ecological interaction between neighbouring
species.  A  general description should include inhomogeneities
in $\epsilon$. Here, for simplicity we limit our analysis to the
homogeneous case, $\epsilon_i \equiv \epsilon$.

We consider this CML as the substrate on which evolution takes
place.  The parameter $\lambda_i$, that determines the behaviour
of $x_i$, is regarded as the microscopic level, the genotype,
and is subject to mutation.  We suppose that, through mutation,
a species is able to alter this strategy to adapt to the
environment defined by the collective behaviour. For this we
propose an {\em extremal} mechanism, akin to the BS model
\cite{BS93}. The species that, at each time step, has the
minimum value of $x$, is considered the candidate to mutation.
Its $\lambda$ is replaced by a new value drawn from the
distribution $g(\lambda)$.

We can summarise these simple dynamical rules by
\begin{enumerate}
\item{} Evaluate expression (\ref{cml}).
\item{} Find the site with the absolute minimum fitness
on the lattice (this site will be called the ``active'' site).
\item{} Change the value of $\lambda$  of the active site.
\item{} Go to step 1.
\end{enumerate}

\begin{figure}[t]
\centerline{\psfig{file=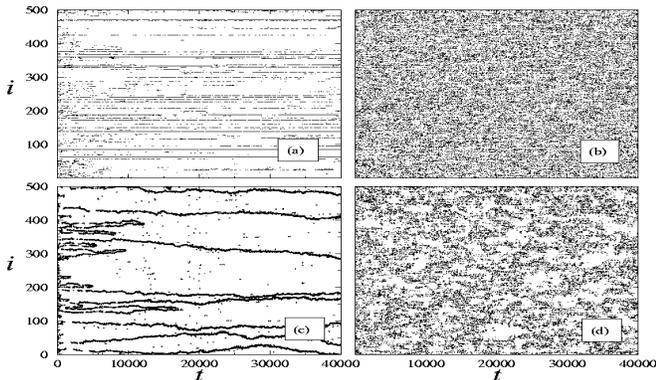,angle=-90,width=\columnwidth,height=5cm}}
\caption{Space-time diagram of the active sites in a CML of
$500$ sites.  Each dot is an active site, i.e. the site with the
minimum $x$ at a certain time. $\Delta\lambda=0.1$ in all four
cases.(a)$\lambda_0=3.7$, $\epsilon=0.5$. (b)$\lambda_0=3.9$,
$\epsilon=0.5$. (c) $\lambda_0=3.8$, $\epsilon=0.1$.
(d)$\lambda_0=3.9$, $\epsilon=0.1$. } \label{active}
\end{figure}

In Fig. \ref{active} we show the space-time picture of the
active sites for four different sets of parameters. In all cases
we take  $g(\lambda)$ uniform within the range
$(\lambda_0,\lambda_0+\Delta\lambda)$ and zero outside, with
$\Delta\lambda = 0.1$. This four cases can be considered
paradigmatic of the behaviours observed in a wide region of
parameter space. Fig. \ref{active}(a) and (b) correspond to a
rather high value of the coupling, $\epsilon=0.5$. Fig.
\ref{active}(c) and (d) have, instead, $\epsilon=0.1$. The value
of $\lambda_0$ is $3.9$ in  (b) and (d), $3.7$ in (a),  and
$3.8$ in (c). Depending on the value of the parameters,  the
behaviour of the successive minima can be classified in one of
the following four categories.
\begin{itemize}
\item[(a)] discontinuous lines at $i=\mbox{constant}$: The activity is concentrated in
a few sites of the system, and in first approximation the same
sites remain active as time goes by.
\item[(b)] uniform: The activity is spread all over the system without any
apparent order.
\item[(c)] ``worms'': The activity is mostly concentrated in
a few sites of the system. The activity, however, wanders as time goes by
following  a random pattern.
\item[(d)] clusterized: The activity is spread all over the system, but in
clusters. At variance with (b), at any given time,  one sees regions in space
where there is no activity.
\end{itemize}

Let us discuss in more detail Fig. \ref{active}(c). The worms
are created during the transient. Once created,  each worm
wanders in space till it encounters a second worm. At that
moment they annihilate each other. We have observed that the
inactive regions correspond to a periodic pattern in space
\cite{Kaneko93}. The worms, in turn, correspond to defects in
the periodicity. Eventually, the last two surviving worms merge,
and from there on the activity remains concentrated on a single
``fat'' worm.

\begin{figure}
\centerline{\psfig{file=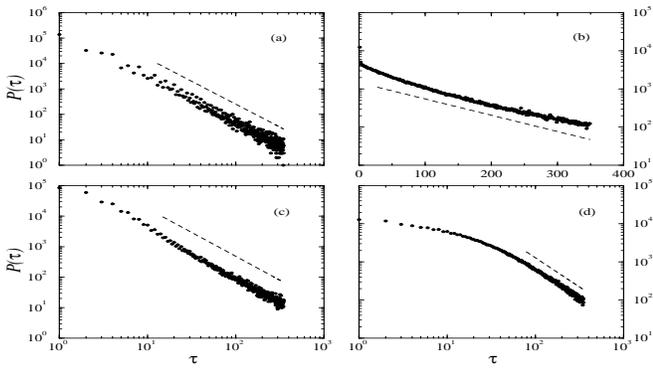,angle=-90,width=\columnwidth,height=5cm}}
\caption{Distribution of first return time. The parameters
correspond to those in Fig. \protect\ref{active}: $N=500$,
$\Delta\lambda=0.1$, (a)$\lambda_0=3.7$, $\epsilon=0.5$. The
dashed line corresponds to a power law with exponent
$\alpha=1.80(3)$. (b)$\lambda_0=3.9$, $\epsilon=0.5$. The dashed
line, shown as a reference, corresponds to an exponential decay.
(c) $\lambda_0=3.8$, $\epsilon=0.1$. The dashed line corresponds
to a power law with exponent  $\alpha=1.57(2)$.
(d)$\lambda_0=3.9$, $\epsilon=0.1$. The dashed line corresponds
to a power law with exponent $\alpha=1.53(1)$, fitted to the
asymptotic region. } \label{firstret}
\end{figure}

For each one of these cases, we have computed the first return
time distribution. This corresponds to the distribution of times
between consecutive activity in the same site. We observe that
when the successive minima are distributed uniformly in space,
the first return time distribution is close to an exponential
(see Fig.\ \ref{firstret}(b)). As soon as the activity is not
uniformly distributed, be it lines or irregular clusters, the
first return time distribution exhibits a power law decay
(asymptotically in case (d)).

\begin{figure}
\centerline{\psfig{file=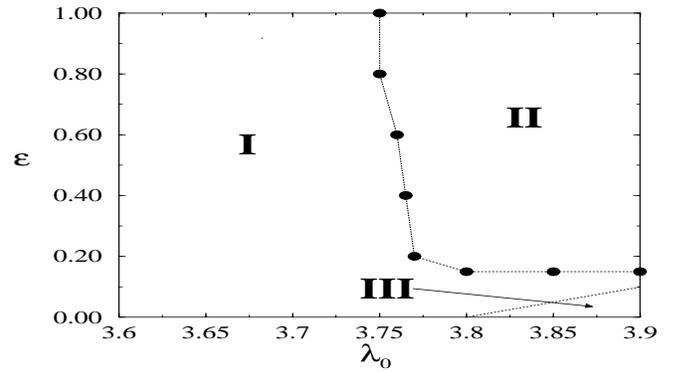,width=\columnwidth,height=5cm}}
\caption{Phase diagram in $\lambda$-$\epsilon$ space. $N=200$,
$\Delta \lambda =0.1$. Averaged over 10 realizations. }
\label{diag}
\end{figure}

To make this picture more quantitative, we present in Fig.\
\ref{diag} a phase diagram, in $\lambda$-$\epsilon$ space
($\Delta \lambda$ is held constant). Phase I is characterised by
the presence of power law behaviour in the  first return time
distribution, $P(\tau)\sim \tau^{-\alpha}$. The value of the
exponent of this distribution is not constant within the region.
Phase II, on the other hand, is characterised by a non power law
behaviour (close to exponential) in the  first return time
distribution (see Fig.\ \ref{firstret}(b)). This is tantamount
to saying that in phase I there is clusterization of the
activity, that is absent in phase II. In Fig.\ \ref{diag} we
indicate also a third phase (III), which corresponds to an
asymptotic power law (see Fig.\ \ref{firstret}(d)), with an
exponent that is different from that in region I. The border of
this region (that we show only schematically) is not  as well
defined as the border between regions I and II.

This picture  holds, qualitatively, even in the absence of the
evolutionary dynamics. Indeed, if in every time step we track
the position of the minimum, but do not change the value of its
$\lambda$ (this corresponds, effectively, to setting
$\Delta\lambda=0$), we still obtain the four abovementioned
regimes and the corresponding first return time distributions.
As  $\Delta\lambda$  changes, the value of the exponent in the
first return time distribution (in region I) changes as well.
Moreover, the area of region II increases with increasing
$\Delta\lambda$.

The presence of evolutionary dynamics (i.\ e.\ $\Delta\lambda >
0$), has yet another effect. The distribution of $\lambda$
evolves in time, from an originally uniform to a stationary
non-uniform one. Indeed, this stationary distribution is peaked
close to $\lambda_0$ and monotonically decresases for larger
values of $\lambda$: The system has ``self-organized''. Since
the extremal dynamics favors higher values of $x$, and the
higher the value of $\lambda$ the more likely small values of
$x$ are \cite{Schuster89}, then large values of $\lambda$ are
more likely to be updated. The exact shape of this distribution
depends on the values of the parameters \cite{AV98}.

Summarizing, the dynamics of extremal coupled maps exhibits a behavior
usually associated with criticality:
\begin{itemize}
\item First return time distribution is a power law (even in the absence of
extremal dynamics)\\
\item there is self-organization in the $\lambda$ space\\
\item the active regions are localized in space, very
much like the avalanches in SOC models (BS).
\end{itemize}

It is worth emphasizing that this behavior (except the self
organization in $\lambda$ space) remains even in the absence of
extremal dynamics. In particular, the first return time
distribution exhibits a power law behavior. This implies that
power  law distributions are compatible  with  non-extremal
systems, as has been previously observed by Newman an coworkers
\cite{Newman}, in the context of noise-driven systems.

Both SOC and CML models have been  used separately in the past
to describe evolving ecosystems. To our knowledge, the model
presented here is the first synthesis of both ideas that
includes the level structure (genotype/phenotype) of real
organisms.



\begin{thebibliography}{99}
\bibitem[\ast]{} E-mail: abramson@cab.cnea.gov.ar. Present
address: Centro At\'{o}mico Bariloche, 8400 Bariloche,
Argentina.
\bibitem[\dagger]{}E-mail: jose@mpipks-dresden.mpg.de.
\bibitem{CJ89} J.M. Carlson and J.S. Langer, Phys. Rev. Lett. {\bf 62}, 2632
(1989).
\bibitem{KS95}L. Knopoff and D. Sornette, J. Physique I {\bf 5}, 1682 (1995).
\bibitem{Zaitsev92} S.I. Zaitsev, Physica A {\bf 189}, 411 (1992).
\bibitem{PPVAC94} A. Petri, G. Paparo, A. Vespignani, A. Alippi and
M. Costantini, Phys. Rev. Lett. {\bf 73}, 3423 (1994).
\bibitem{DMP91} P. Diodati, F. Marchesoni and S. Piazza, Phys. Rev. Lett. {\bf
67}, 2239 (1991).
\bibitem{CTP96} G. Caldarelli, F. di Tolla and A. Petri, Phys. Rev. Lett. {\bf
77}, 2503 (1996).
\bibitem{WW83} D. Wilkinson and J.F. Willemsen, J. Phys. A {\bf 16}, 3365
(1983).
\bibitem{CR88} M. Cieplak and M.O. Robbins, Phys. Rev. Lett. {\bf 60}, 2042
(1988).
\bibitem{Sneppen92} K. Sneppen, Phys. Rev. Lett. {\bf 69}, 3539 (1992).
\bibitem{Sneppen93} K. Sneppen, Phys. Rev. Lett. {\bf 71}, 101 (1993).
\bibitem{RRRIB93} A. Rinaldo, I. Rodriguez-Iturbe, R. Rigon, E. Ijjasz-Vasquez
and R.L. Bras, Phys. Rev. Lett. {\bf 70}, 822 (1993).
\bibitem{MCFCB96} A. Maritan, F. Colaiori, A. Flammini, M. Cieplak and
J.R. Banavar, Science {\bf 272}, 984 (1996).
\bibitem{CGMRR96} G. Caldarelli, A. Giacometti, A. Maritan,
I. Rodriguez-Iturbe and A. Rinaldo, Phys. Rev. E {\bf 55}, R4865 (1997).
\bibitem{Raup86} M.D. Raup, Science {\bf 251}, 1530 (1986).
\bibitem{EG72} N. Eldredge and S.J. Gould, {\it Punctuated equilibria: an
alternative to phyletic gradualism} in {\it Models in Paleobiology}, Freeman,
Cooper and Co., San Francisco (1972).
\bibitem{GE77} S.J. Gould and N. Eldredge, Paleobiology {\bf 3}, 115 (1977).
\bibitem{EG88} N. Eldredge and S.J. Gould, Nature {\bf 332}, 211 (1988).
\bibitem{BTW87} P. Bak, C. Tang and K. Wiesenfeld, Phys. Rev. Lett. {\bf
59}, 381 (1987).
\bibitem{BTW88} P. Bak, C. Tang and K. Wiesenfeld, Phys. Rev. A {\bf 38}, 364
(1988).
\bibitem{PMB96} M. Paczuski, S. Maslov and P. Bak, Phys. Rev. E {\bf 53}, 414
(1996).
\bibitem{BS93} P. Bak and K. Sneppen, Phys. Rev. Lett. {\bf 71}, 4083 (1993).
\bibitem{DVV97} P. De Los Rios, A. Valleriani and J. L. Vega,  Phys. Rev. E
 {\bf 56}, 4876 (1997).
\bibitem{DVV98} P. De Los Rios, A. Valleriani and J. L. Vega,  Phys. Rev. E
 {\bf 57}, 6451 (1998).
\bibitem{MPB94} S. Maslov, M. Paczuski and P. Bak, Phys. Rev. Lett. {\bf 73},
2162 (1994).
\bibitem{Grassberger95} P. Grassberger, Phys. Lett. A {\bf 200}, 277 (1995).

\bibitem{comment2} The logistic map is a paradigmatic model of the evolution of $x_i$
in the context of population dynamics. As such, it has been carefully studied
and many its properties are well known.
\bibitem{comment1} Coupled Map Lattices have been extensively
studied by Kaneko and coworkers \cite{Kaneko93},   as well as many
other authors (see \cite{CM92} and references therein). They exhibit
  frozen patterns as well as
travelling waves depending on the values of $\epsilon$ and
$g(\lambda)$.
\bibitem{Kaneko93}K. Kaneko, {\em Theory and applications of coupled
map lattices} (John Wiley \& Sons, Chichester, 1993).
\bibitem{CM92} H. Chat\'{e} and P. Manneville, Prog. Theor. Phys. {\bf 87}, 1 (1992).
\bibitem{Schuster89} H. G. Schuster, {\it Determistic Chaos} VCH
Verlag, Weinheim, 1989).
\bibitem{AV98} This topic is currently under investigation.
\bibitem{Newman} B. W. Roberts and M. E. J. Newman, J. Theor. Biol. {\bf 180},
39 (1996).
\end{thebibliography}
\end{document}